\documentclass[10pt,conference]{IEEEtran}

\IEEEoverridecommandlockouts

\usepackage{cite}
\usepackage{amsmath,amssymb,amsfonts}
\usepackage{algorithmic}
\usepackage{graphicx}
\usepackage{textcomp}
\usepackage{xcolor}
\usepackage{amsthm}
\usepackage{amssymb}
\usepackage{amsmath}
\usepackage{subfigure}
\usepackage[linesnumbered, ruled]{algorithm2e}

\def\BibTeX{{\rm B\kern-.05em{\sc i\kern-.025em b}\kern-.08em
    T\kern-.1667em\lower.7ex\hbox{E}\kern-.125emX}}
\usepackage{geometry}
\begin{document}
\title{Simultaneous Beamforming and Anti-Jamming With Intelligent Omni-Surfaces\\
}
\author{
Yuhan Wang$^{1,2}$, Shuhao Zeng$^{1,4}$, Qingyu Liu$^{1}$, Boya Di$^{3}$, Hongliang Zhang$^{3}$\\
$\quad^{1}$School of Electronic and Computer Engineering, Peking University Shenzhen Graduate School, China \\
$\quad^{2}$Pengcheng Laboratory, China\\
$\quad^{3}$School of Electronics, Peking University, China\\
$\quad^{4}$Department of Electrical and Computer Engineering, Princeton University, USA  \\
\vspace{-1cm}}

\maketitle

\begin{abstract}
Wireless transmission is vulnerable to malicious jamming attacks due to the openness of wireless channels, posing a severe threat to wireless communications. Current anti-jamming studies primarily focus on either enhancing desired signals or mitigating jamming, resulting in limited performance. To address this issue, intelligent omni-surface (IOS) is a promising solution. By jointly designing its reflective and refractive properties, the IOS can simultaneously nullify jamming and enhance desired signals. In this paper, we consider an IOS-aided multi-user anti-jamming communication system, aiming to improve desired signals and nullify jamming by optimizing IOS phase shifts and transmit beamforming. However, this is challenging due to the coupled and discrete IOS reflection and refraction phase shifts, the unknown jammer's beamformer, and imperfect jammer-related channel state information. To tackle this, we relax IOS phase shifts to continuous states and optimize with a coupling-aware algorithm using the Cauchy-Schwarz inequality and S-procedure, followed by a local search to recover discrete states. Simulation results show that the proposed scheme significantly improves the sum rate amid jamming attacks.
\end{abstract}

\vspace{-0.2cm}
\section{Introduction}
Due to the broadcast and openness of wireless channels~\cite{wu2022Hybrid,lu2022Integrated}, wireless transmissions are susceptible to jamming attacks, which reduce signal-to-interference-plus-noise ratio (SINR) and degrade communication capacity. Recently, reconfigurable intelligent surfaces (RISs) have been utilized for anti-jamming due to the capability of reshaping the propagation environment~\cite{ElMossallamy2020Reconfigurable}. Specifically, an RIS is an ultra-thin surface, which can apply adjustable phase shifts to incident signals to reshape the transmission environment as desired~\cite{zeng2021Reconfigurable}, thereby improving SINR. However, most existing research focused on reflective-only RIS, known as intelligent reflecting surface (IRS), serving users on one side only and making it challenging to align desired beams with the null of jamming beams simultaneously, thus limiting RIS-assisted anti-jamming performance. 

\begin{figure}[!t]
	\centering
	\includegraphics[width=0.22\textwidth]{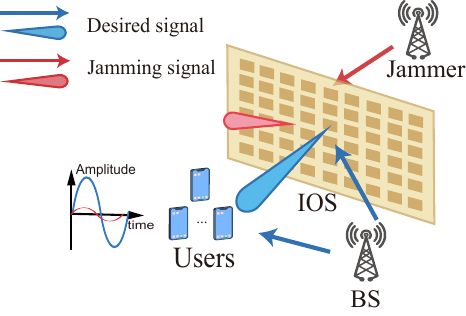}
	\vspace{-3mm}
	\caption{System model of IOS-aided anti-jamming multiuser communications. (Take users on the same side as the BS as an example.)}
	
	\label{sysmodel}
\end{figure}

To cope with this issue, intelligent omni-surface (IOS) is a promising solution, which is capable of simultaneous signal reflection and refraction~\cite{zhang2022intelligent}. By adjusting the ON/OFF state of the positive-intrinsic-negative (PIN) diodes loaded in each element, the reflected and refracted phase modulation of the impinging signal is achieved~\cite{zeng2022intelligent}. Therefore, the IOS can nullify jamming signals and enhance desired signals for users on both sides by applying reconfigurable reflection and refraction phase shifts, respectively. To illustrate the working principle, we take the user on the same side as the base station (BS) as an example, as shown in Fig.~\ref{sysmodel}. By adjusting the IOS reflection and refraction phase shifts, the desired signal is reflected by the IOS towards the user and coherently superposed with those from the BS-user direct link to enhance the received power. Meanwhile, the jamming signal is refracted by the IOS and counteracts those from the jammer-user direct link, making the null of the jamming beam towards the user. 

In this paper, we consider an IOS-aided multi-user anti-jamming communication system. We aim to reduce jamming and enhance desired signals by jointly optimizing BS digital beamforming and IOS phase shifts. This is challenging due to the coupled and discrete IOS phase shifts and the lack of coordination between the BS and the jammer, resulting in the unknown jammer's beamformer and imperfect jammer-related channel state information (CSI), both crucial for effective beamforming. To address the challenge, we relax phase shifts to continuous states and develop a coupling-aware algorithm with the Cauchy-Schwarz inequality and the S-procedure, followed by a local search to recover discrete states.

In the literature, most existing studies on RIS-assisted anti-jamming communications focus on IRSs, divided based on the role of the IRS into those that enhance desired signals~\cite{yang2020intelligent} and those that consider mitigating jamming signals to meet system requirements~\cite{sun2021intelligent,sun2022outage}. Specifically, authors in~\cite{yang2020intelligent} proposed a reinforcement learning-based approach for IRS-assist multiuser anti-jamming, and authors in~\cite{sun2022outage} studied robust beamforming designs minimizing BS transmit power while ensuring SINR requirements. There are few studies on IOS-aided anti-jamming. In~\cite{zhou2023robust}, the authors used an IOS to cover users on both sides and designed robust beamforming based on continuous and independent reflection and refraction phase shifts. 
Unlike these works, we apply IOS reflection and refraction to achieve simultaneous jamming cancellation and desired signal enhancement, considering a more practical assumption of discrete and coupled reflection and refraction phase shifts of IOS. Existing beamforming designs are inapplicable, we need to consider the coupling effects between IOS reflection and refraction.

\textit{Contributions:} We propose an IOS-aided anti-jamming scheme that simultaneously nullifies jamming and enhances desired signals by applying reconfigurable reflection and refraction phase shifts, respectively. We formulate a joint BS digital beamforming and IOS analog beamforming optimization problem, which is decomposed and solved alternately. We adopt zero-forcing beamforming for digital beamforming and relax the discrete IOS analog beamforming problem into a continuous one, solving by a coupling-aware algorithm with the Cauchy-Schwarz inequality and S-procedure, followed by a local search to recover discrete states. Simulations show that the proposed scheme effectively nullifies jamming and enhances desired signals, and achieves a higher sum rate than IRS-aided anti-jamming schemes. 

\textit{Organization of the rest part:} Section~II details the considered multi-user IOS-aided anti-jamming communication system and presents the joint optimization problem and its decomposition. Section~III elaborates on the algorithm developed to solve the decomposed problems. Section~IV displays the results of the simulation to assess the performance of the proposed scheme. Finally, conclusions are given in Section~V.

\vspace{-0.1cm}
\section{System Model and Problem Formulation}
\subsection{Scenario Description}
As shown in Fig.~\ref{sysmodel}, we consider a downlink multi-user communication system where a BS equipped with $N_b$ antennas transmits to $K$ single-antenna users, denoted by $\mathcal{K}=\{1,2,...K\}$. A jammer with $N_j$ antennas emits the jamming signal. To cope with it, we deploy an IOS between the BS and the jammer, and utilize its simultaneous reflection and refraction capability to enhance the desired signal and nullify the jamming signal at the same time. The users are split into two subsets based on their location relative to the IOS: $\mathcal{K}^R$ for users on the same side as the BS, with $K_R$ users in it; $\mathcal{K}^T$ for users on the opposite side, with $K_T$ users in it.

\subsection{IOS Model}
An IOS consists of $M$ reconfigurable elements denoted by $\mathcal{M}=\{1,2,...M\}$. As shown in Fig.~\ref{sysmodel}, each element is equipped with multiple PIN diodes that switch between ON and OFF states based on applied voltages, determining the state of each element. When a signal impinges on the IOS element, it is simultaneously reflected and refracted with phase shifts adjusted by the element state~\cite{zhang2021intelligent}. The reflection and refraction phase shifts of the IOS element are coupled, meaning both reflection and refraction phase shifts change together when adjusting the element state. Assume each element can offer $2^b$ different states, the set of available reflection and refraction phase shifts of the $m$-th element can be given by
\begin{align}
\label{phaseshiftset}
  (\phi_m^r,\phi_m^t)\in\mathcal{F}=\{(\varphi_1^r,\varphi_1^t),(\varphi_2^r,\varphi_2^t),...,(\varphi_{2^b}^r,\varphi_{2^b}^t)\}.  
\end{align}
Denote the reflection and refraction responses of the $m$-th element of the IOS to the incident signal by $g_{m}^{r}$ and $g_{m}^{t}$, respectively, which can be modeled by~\cite{zeng2022intelligent}
\begin{align}
\label{element_response1}
g_{m}^{r}=\Gamma_m^re^{-j\phi_{m}^{r}}, g_{m}^{t}=\Gamma_m^te^{-j\phi_{m}^{t}}.
\end{align}
Here, $\Gamma_m^r$ and $\Gamma_m^t$ denote the amplitude of the reflection and refraction responses, respectively. Based on (\ref{element_response1}), the reflection response of the IOS to the signal can be given by 
\begin{align}
\label{qr}
\mathbf{g}_{r}=[g_1^r, g_2^r, ..., g_M^r]^H=\mathbf{A}_{r}\mathbf{q}_{r},
\end{align}
where amplitute matrix $\mathbf{A}_{r}=\text{diag}([\Gamma_1^r, \Gamma_2^r, ..., \Gamma_M^r]^H)$ and phase shift vector $\mathbf{q}_{r}=[e^{-j\phi_{1}^{r}}, e^{-j\phi_{2}^{r}}, ..., e^{-j\phi_{M}^{r}}]^H$. The refraction response of the IOS to the signal is given by 
\begin{align}
\label{qt}
\mathbf{g}_{t}=[g_1^t, g_2^t, ..., g_M^t]^H=\mathbf{A}_{t}\mathbf{q}_{t},
\end{align}
where  amplitute matrix $\mathbf{A}_{t}=\text{diag}([\Gamma_1^t, \Gamma_2^t, ..., \Gamma_M^t]^H)$ and phase shift vector $\mathbf{q}_{t}=[e^{-j\phi_{1}^{t}}, e^{-j\phi_{2}^{t}}, ..., e^{-j\phi_{M}^{t}}]^H$.

\subsection{Channel Model}
In this section, we assume the Racian model for data channels from the BS to user $k$ and jamming channels from the jammer to the user $k$.

The data channels from the BS to user $k$ primarily consist of the reflected/refracted channels via the IOS, denoted by $\mathbf{H}_{B,k}\in\mathbb{C}^{M\times N_b}$~\footnote{For ease of analysis, we separate the IOS response from the channel $\mathbf{H}_{B,k}$ and represent it using (\ref{qr}) and (\ref{qt}). In reality, the channel between the BS and user $k$ via IOS includes the IOS response.}, and the direct link, denoted by $\mathbf{h}_{B,k}\in\mathbb{C}^{N_b\times 1}$. Each element in $\mathbf{H}_{B,k}$ and $\mathbf{h}_{B,k}$ is modeled as a Rician model, where the dominant LoS component is formed by the BS-IOS-user link and BS-user link, respectively, and the non-LoS (NLoS) component is constituted by all other paths~\cite{di2020hybrid}. For example, the $(m,n_b)$-th element ${h}_{B,k,(m,n_b)}$ in $\mathbf{H}_{B,k}$ is given by

\vspace{-0.3cm}
\begin{small}
\begin{equation}
  \begin{aligned}
\label{HBK}
{h}_{B,k,(m,n_b)}=\sqrt{\frac{\kappa}{1+\kappa}}{h}_{B,k,(m,n_b)}^{LoS}+\sqrt{\frac{1}{1+\kappa}}{h}_{B,k,(m,n_b)}^{NLoS}.
\end{aligned}  
\end{equation}
\end{small}
Here, $\kappa$ denotes the Rician factor. ${h}_{B,k,(m,n_b)}^{LoS}$ is the LoS component, which can be expressed as~\cite{zhang2021intelligent}
\begin{small}
\begin{equation}
\begin{aligned}
\label{h_nb_k_m_los}
{h}_{B,k,(m,n_b)}^{LoS}&=\frac{\lambda\sqrt{G_{n_b}^{tx}K^A(m)G_{k}^{rx}K_k^D(m)}e^{-j\frac{2\pi}{\lambda}(d_{n_b,m}+d_{k,m})}}{(4\pi)^{\frac{3}{2}}d_{n_b,m}^{\alpha}d_{k,m}^{\alpha}}.
\end{aligned}
\end{equation}
\end{small}
Here, $\lambda$ is the signal wavelength, $G_{n_b}^{tx}$ and $G_{k}^{rx}$ are the power gains of the $n_b$-th antenna of the BS and the antenna of user $k$, respectively. $K^A(m)$ and $K_k^D(m)$ are the normalized power gain of the $n_b$-th antenna of the BS and user $k$ in the direction of the $m$-th element of the IOS, respectively. $d_{n_b,m}$ and $d_{k,m}$ are the transmission distances between the $m$-th element of the IOS and the $n_b$-th antenna of the BS and user $k$, respectively, and $\alpha$ is the corresponding path-loss exponent. 
The NLoS component ${h}_{B,k,(m,n_b)}^{NLoS}$ is expressed as~\cite{di2020hybrid}
\begin{align}
\label{h_nb_k_m_NLOS}
{h}_{B,k,(m,n_b)}^{NLoS}=PL(d_{n_b,m},d_{k,m})h^{SS},
\end{align}
where $PL(\cdot)$ is the path loss model for NLoS transmissions, and $h^{SS} \sim CN (0,1)$ accounts for the cumulative effect of multiple scattered paths that originate from the random scatterers available in the propagation environment.

The $n_b$-th element in $\mathbf{h}_{B,k}$ is also modeled as a Rician model. 
The LoS component ${h}_{B,k,(n_b)}^{direct,LoS}$ is given by~\cite{zhang2021intelligent}
\begin{align}
\label{h_nb_k_los}
{h}_{B,k,(n_b)}^{direct,LoS}=\sqrt{G_{n_b}^{tx}F_{n_b,k}G_{k}^{rx}d_{n_b,k}^{-\alpha}}e^{-j\frac{2\pi}{\lambda}d_{n_b,k}},
\end{align}
where $F_{n_b,k}$ is the normalized end-to-end power gain of the $n_b$-th antenna of the BS and user $k$. $d_{n_b,k}$ is the distance between the $n_b$-th antenna of the BS and user $k$. 
The NLoS component ${h}_{B,k,(n_b)}^{direct,NLoS}$ is expressed by~\cite{di2020hybrid}
\begin{align}
\label{h_nb_k_nlos}
{h}_{B,k,(n_b)}^{direct,NLoS}=PL(d_{n_b,k})h^{SS}.
\end{align}

Similarly, the jamming channels from the jammer to user $k$ consist of the reflected/refracted channels via the IOS, denoted 
by $\mathbf{H}_{J,k}\in\mathbb{C}^{M\times N_j}$, and the direct link, denoted 
by $\mathbf{h}_{J,k}\in\mathbb{C}^{N_j\times 1}$. Similar to the data channels, the jamming channels (i.e., $\mathbf{h}_{J,k}$ and $\mathbf{H}_{J,k}$) are modeled using the aforementioned method. 
However, due to the absence of collaboration between the BS and the jammer, it is hard to acquire the jamming channels accurately, leading to so-called channel uncertainty~\cite{sun2021intelligent}. To describe the uncertainty of the jamming channels, the bounded CSI model is adopted, which can be represented as
\vspace{-0.1cm}
\begin{align}
\label{h_JK1}
\mathbf{h}_{J,k}={\hat{\mathbf{h}}}_{J,k}+\Delta\mathbf{h}_{J,k},\  \mathbf{H}_{J,k}={\hat{\mathbf{H}}}_{J,k}+\Delta\mathbf{H}_{J,k},
\end{align}
where $\hat{\mathbf{h}}_{J,k}$ and $\hat{\mathbf{H}}_{J,k}$ are estimated channels for the direct channel $\mathbf{h}_{J,k}$ and the IOS scattering channel $\mathbf{H}_{J,k}$ , respectively. $\Delta\mathbf{h}_{J,k}$ and $\Delta\mathbf{H}_{J,k}$ are corresponding channel estimation errors, which are bounded by $\epsilon_{d,k}$ and $\epsilon_{J,k}$, i.e.,
\begin{align}
\label{hJKerror}
\Vert\triangle \mathbf{h}_{J,k}\Vert_2\le\epsilon_{d,k}, \Vert\triangle \mathbf{H}_{J,k}\Vert_F\le\epsilon_{J,k}.
\end{align}

\subsection{Signal Model}
The BS utilizes a digital beamformer $\mathbf{V}_B\in\mathbb{C}^{N_b\times K}$ (with $N_b\geq K$) to encode $K$ data streams $\mathbf{x}\in\mathbb{C}^{K\times1}$, and then sends signals via $N_b$ antennas. A jammer disrupts the communications by sending jamming $\mathbf{v}_J\mathbf{s}_J\in\mathbb{C}^{N_j\times1}$, where $\mathbf{v}_J$ is the jammer's beamformer and $\mathbf{s}_J$ is the jamming signal. The received signal at user $k$ can be given by (\ref{received_signal}) shown at the top of the next page, where $\mathbf{v}_{B,k}$ represents the $k$-th column of $\mathbf{V}_B$, and ${x}_{k}$ is the $k$-th element of $\mathbf{x}$. The term ${n}_k\sim CN(0,\sigma^2)$ denotes the additive white Gaussian noise (AWGN) with zero mean and variance $\sigma^2$. The received SINR at user $k$ can be given by~\cite{zhang2021intelligent}
\begin{figure*}[!t]
\normalsize
\setcounter{equation}{\value{equation}}
\begin{small}
\begin{align}
\label{received_signal}
    y_k=\left\{
    \begin{aligned}
        &           (\mathbf{h}_{B,k}^H+\mathbf{g}_{r}^H\mathbf{H}_{B,k})\mathbf{v}_{B,k}{x}_{k}+\sum_{k\prime\neq k}(\mathbf{h}_{B,k}^H+\mathbf{g}_{r}^H\mathbf{H}_{B,k})\mathbf{v}_{B,k\prime}{x}_{k\prime}+(\mathbf{h}_{J,k}^H+\mathbf{g}_{t}^H\mathbf{H}_{J,k})\mathbf{v}_J\mathbf{s}_{J}+n_k, &&k\in\mathcal{K}^R,\\
        &(\mathbf{h}_{B,k}^H+\mathbf{g}_{t}^H\mathbf{H}_{B,k})\mathbf{v}_{B,k}{x}_{k}+\sum_{k\prime\neq k}(\mathbf{h}_{B,k}^H+\mathbf{g}_{t}^H\mathbf{H}_{B,k})\mathbf{v}_{B,k\prime}{x}_{k\prime}+(\mathbf{h}_{J,k}^H+\mathbf{g}_{r}^H\mathbf{H}_{J,k})\mathbf{v}_J\mathbf{s}_{J}+n_k, &&k\in\mathcal{K}^T.
    \end{aligned}
    \right.
\end{align}
\end{small}
\hrulefill
\vspace{-0.5cm}
\end{figure*}

\vspace{-0.2cm}
\begin{small}
\begin{align}
\label{SINR}
    \text{SINR}_k=\frac{\left|(\mathbf{h}_{B,k}^H+\mathbf{g}_{s}^H\mathbf{H}_{B,k})\mathbf{v}_{B,k}\right|^2}{\sum_{k\prime\neq k}\left|(\mathbf{h}_{B,k}^H+\mathbf{g}_{s}^H\mathbf{H}_{B,k})\mathbf{v}_{B,k\prime}\right|^2+J_k+\sigma^2},
\end{align}
\end{small}
where $\mathbf{g}_{s}=\mathbf{g}_{r},k\in\mathcal{K}^R$ and $\mathbf{g}_{s}=\mathbf{g}_{t},k\in\mathcal{K}^T$. $J_k$ denotes the received jamming power at user $k$, which is given by

\vspace{-0.4cm}
  \begin{align}
 \label{J_k}
     J_k=\left|(\mathbf{h}_{J,k}^H+\mathbf{g}_{s}^H\mathbf{H}_{J,k})\mathbf{v}_J\right|^2,
 \end{align}
where $\mathbf{g}_{s}=\mathbf{g}_{t},k\in\mathcal{K}^R$ and $\mathbf{g}_{s}=\mathbf{g}_{r},k\in\mathcal{K}^T$. Based on the received SINR given in (\ref{SINR}), the achievable rate for user $k$ can then be given by $R_k=\log_2{(1+\text{SINR}_k)}$.
 
\subsection{Problem Formulation}
In this paper, our goal is to maximize the sum rate by jointly designing the BS precoder $\mathbf{V}_{B}$ and the discrete phase shifts ${\phi_m^r}$ and ${\phi_m^t}$, ensuring the jamming power received by each user below a predetermined threshold. The optimization problem is formally formulated as

\vspace{-0.5cm}
 \begin{subequations}\label{sequestion}
 \begin{align}
 \label{sequestiona}
 &\max_{\mathbf{V}_{B},\{\phi_m^r\},\{\phi_m^t\}}{\sum_{k=1}^{K}R_k},\\
 \label{sequestionb}
 ~\text{s.t.}\ &\text{Tr}(\mathbf{V}_{B}\mathbf{V}_{B}^H)\le P_T,\\
 \label{sequestionc}
 & (\phi_m^r,\phi_m^t)\in\mathcal{F},\ \forall m\in\mathcal{M}, \\
     \label{sequestionf}
     &J_k\le\tau_k, \forall k\in\mathcal{K}, \Vert\triangle \mathbf{h}_{J,k}\Vert_2\le\epsilon_{d,k}, \Vert\triangle \mathbf{H}_{J,k}\Vert_F\le\epsilon_{J,k}.
 \end{align}
 \end{subequations}
 Constraint (\ref{sequestionb}) limits the transmit power below the maximum power $P_T$. Constraint (\ref{sequestionc}) states that the reflection and refraction phase shifts of the IOS element are selected from the set $\mathcal{F}$ provided in (\ref{phaseshiftset}). Constraint (\ref{sequestionf}) restricts the received jamming power by user $k$ below the threshold $\tau_k$.

 To effectively address the problem, we decompose (\ref{sequestion}) into two subproblems as shown below.
 \subsubsection{Digital Beamforming Subproblem at the BS}
Given IOS phase shifts $\{\phi_m^r\}$ and $\{\phi_m^t\}$, the digital beamforming subproblem can be written by
\vspace{-0.1cm}
 \begin{subequations}\label{vbquestion}
 \begin{align}
 \label{vbquestiona}
 &\max_{\mathbf{V}_{B}}{\sum_{k=1}^{K}R_k},\\
 \label{vbquestionb}
 ~\text{s.t.}\ &\text{Tr}(\mathbf{V}_{B}\mathbf{V}_{B}^H)\le P_T.
 \end{align}
 \end{subequations}
 
 \subsubsection{IOS-based Analog Beamforming Subproblem}
 The IOS-based analog beamforming subproblem with fixed beamformer $\mathbf{V}_{B}$ is equivalent to
 \vspace{-0.2cm}
  \begin{subequations}\label{qquestion}
 \begin{align}
 \label{qquestiona}
 &\max_{\{\phi_m^r\},\{\phi_m^t\}}{\sum_{k=1}^{K}R_k},\\
 \label{qqquestionb}
  ~\text{s.t.}\ & (\phi_m^r,\phi_m^t)\in\mathcal{F},\ \forall m\in\mathcal{M}, \\
     \label{qquestione}
     &J_k\le\tau_k, \forall k\in\mathcal{K}, \Vert\triangle \mathbf{h}_{J,k}\Vert_2\le\epsilon_{d,k}, \Vert\triangle \mathbf{H}_{J,k}\Vert_F\le\epsilon_{J,k}.
 \end{align}
 \end{subequations}

\section{Simultaneous jamming cancellation and desired signal enhancement: Algorithm Design}
We first develop algorithms to individually address subproblems (\ref{vbquestion}) and (\ref{qquestion}), respectively. Subsequently, an iterative algorithm is proposed to solve the overall problem (\ref{sequestion}).

\subsection{Digital Beamforming Optimization at the BS}
  \label{vbopt}
 To address the digital beamforming problem in (\ref{vbquestion}), we adopt zero-forcing (ZF) beamforming at the BS to mitigate inter-user interference. We define $\mathbf{H}\in\mathbb{C}^{K\times N_b}$ as the channel matrix of the desired signal transmitted from the BS to $K$ users. The $k$-th row of $\mathbf{H}$ is the channel between the BS and user $k$, denoted by $\mathbf{h}_k^H$, which is given by
 \begin{align}
 \mathbf{h}_k=(\mathbf{h}_{B,k}^H+\mathbf{g}_{s}^H\mathbf{H}_{B,k})^H,\forall k\in\mathcal{K},
 \end{align} 
 where $\mathbf{g}_{s}=\mathbf{g}_{r},k\in\mathcal{K}^R$ and $\mathbf{g}_{s}=\mathbf{g}_{t},k\in\mathcal{K}^T$. 
 Then the ZF beamforming can be expressed as~\cite{di2020hybrid,guo2022Prospects}
 \begin{align}
\label{VBZF}
\mathbf{V}_B=\mathbf{H}^{H}(\mathbf{H}\mathbf{H}^{H})^{-1}\mathbf{P}^{\frac{1}{2}}=\widetilde{\mathbf{V}}_B\mathbf{P}^{\frac{1}{2}},
\end{align}
where $\widetilde{\mathbf{V}}_B=\mathbf{H}^{H}(\mathbf{H}\mathbf{H}^{H})^{-1}$. $\mathbf{P}$ is a diagonal matrix whose $k$-th diagonal element $p_k$ is the received power at the $k$-th user. Based on the digital beamformer in (\ref{VBZF}), problem (\ref{vbquestion}) can be rewritten as
\vspace{-0.2cm}
\begin{subequations}\label{vbquestion1}
 \begin{align}
 \label{vbquestiona1}
 &\max_{p_k\geq0}{\sum_{k=1}^{K}\log_2{\left(1+\frac{p_k}{J_k+\sigma^2}\right)}},\\
 \label{vbquestionb1}
 ~\text{s.t.}\ &\text{Tr}(\mathbf{P}^{1/2}{\widetilde{\mathbf{V}}}_B^H{\widetilde{\mathbf{V}}}_B\mathbf{P}^{1/2})\le P_T.
 \end{align}
 \end{subequations}
The optimal solution of problem (\ref{vbquestion1}) can be obtained by water-filling as $p_k=\frac{1}{\nu_k}\max{(\frac{1}{\mu}}-\nu_k(J_k+\sigma^2),0)$~\cite{di2020hybrid}, 
where $\nu_k$ is the $k$-th diagonal element of ${\widetilde{\mathbf{V}}}_B^H{\widetilde{\mathbf{V}}}_B$ and $\mu$ is a normalization factor that fulfills the constraint $\sum_{k=1}^{K}{\max{(\frac{1}{\mu}}-\nu_k(J_k+\sigma^2),0)=P_T}$. 
After obtaining $\mathbf{P}$, the digital beamforming matrix $\mathbf{V}_B$ can be obtained from (\ref{VBZF}).

\subsection{Analog Beamforming Optimization at the IOS}\label{FAT}

In this section, we solve the IOS analog beamforming stated in (\ref{qquestion}), which is difficult to solve since $\{\phi_m^r\}$ and $\{\phi_m^t\}$ are discrete and coupled. We adopt the following three steps: 

\subsubsection{Step 1 - Problem relaxation} 
First, we relax the problem by considering continuous phase shifts. To describe the coupling between the reflection and refraction, we perform a linear fitting on the discrete phase shifts in $\mathcal{F}$, obtaining the constraint (\ref{qquestionb2}). And due to the BS adopting ZF beamforming, the problem (\ref{qquestion}) can be written as
  \begin{subequations}\label{qquestion2}
 \begin{align}
 \label{qquestiona2}
 &\max_{\{\phi_m^r\},\{\phi_m^t\}}{\sum_{k=1}^{K}\log_2{\left(1+\frac{\left|\mathbf{h}_k^H\mathbf{v}_{B,k}\right|^2}{J_k+\sigma^2}\right)}},\\
 \label{qquestionb2}
 ~\text{s.t.}\ 
    &\phi_m^t=s\phi_m^r+v,\forall m\in\mathcal{M},\\
      \label{qquestionc2}
     &J_k\le\tau_k, \forall k\in\mathcal{K}, \Vert\triangle \mathbf{h}_{J,k}\Vert_2\le\epsilon_{d,k}, \Vert\triangle \mathbf{H}_{J,k}\Vert_F\le\epsilon_{J,k}.
 \end{align}
 \end{subequations}

\subsubsection{Step 2 - Continuous IOS Phase Shift Design}
 However, it is challenging to solve problem (\ref{qquestion2}) because it involves unknown $\mathbf{v}_J$ and channel uncertainties. To solve these challenges, we adopt the following methods.
 
 \textit{i) Jammer's beamformer elimination:}
First, the unknown $\mathbf{v}_J$ can be eliminated by using the Cauchy-Schwarz inequality to obtain an upper bound on the jamming power $J_k$ received by the user, which is independent of $\mathbf{v}_J$~\cite{sun2022outage}:

\vspace{-0.4cm}
\begin{small}
\begin{equation}
\begin{aligned}
\label{C5a1}
J_k=\left|(\mathbf{h}_{J,k}^H+\mathbf{g}_{s}^H\mathbf{H}_{J,k})\mathbf{v}_J\right|^2\le P_J\left\|(\mathbf{h}_{J,k}^H+\mathbf{g}_{s}^H\mathbf{H}_{J,k})\right\|_2^2,
\end{aligned}
\end{equation}
\end{small}
where $\mathbf{g}_{s}=\mathbf{g}_{t},k\in\mathcal{K}^R$ and $\mathbf{g}_{s}=\mathbf{g}_{r},k\in\mathcal{K}^T$. Here, $P_J=\left\|\mathbf{v}_J\right\|_2^2$. However, $P_J$ cannot be precisely obtained due to jamming channel uncertainties~\cite{sun2021intelligent}. We can obtain the estimated value $\hat{P}_J$, where the estimation error between $P_J$ and $\hat{P}_J$ is bounded by $\varepsilon_{PJ}$, i.e., $\left|P_J-\hat{P}_J\right|/P_J \le \varepsilon_{PJ}$. By substituting (\ref{C5a1}) to problem (\ref{qquestion2}), $\mathbf{v}_J$ can be eliminated.

\textit{ii) Problem transformation to SDP form:} Next, note that the variables $\{\phi_m^r\}$ and $\{\phi_m^t\}$ in problem (\ref{qquestion2}) appear in absolute value squared terms and Euclidean norm squared terms. Therefore, problem (\ref{qquestion2}) can be transformed into a more tractable SDP form. 
We define $\mathbf{\Xi}_r={\widetilde{\mathbf{q}}}_{r}{\widetilde{\mathbf{q}}}_{r}^{H}$, ${\widetilde{\mathbf{q}}}_{r}=[\mathbf{q}_{r}^{H},1]^H$, $\mathbf{\Xi}_t={\widetilde{\mathbf{q}}}_{t}{\widetilde{\mathbf{q}}}_{t}^{H}$, ${\widetilde{\mathbf{q}}}_{t}=[\mathbf{q}_{t}^{H},1]^H$, $\mathbf{F}_{B,k}=[\mathbf{H}_{B,k}^{H}\mathbf{A}_r,\mathbf{h}_{B,k}], \forall k\in\mathcal{K}^R$, $\mathbf{F}_{B,k}=[\mathbf{H}_{B,k}^{H}\mathbf{A}_t,\mathbf{h}_{B,k}], \forall k\in\mathcal{K}^T$, $\mathbf{V}_{B,k}=\mathbf{v}_{B,k}\mathbf{v}_{B,k}^H$, and 
 $\mathbf{F}_{J,k}=[\mathbf{H}_{J,k}^{H}\mathbf{A}_t,\mathbf{h}_{J,k}],\forall k\in\mathcal{K}^R$, $\mathbf{F}_{J,k}=[\mathbf{H}_{J,k}^{H}\mathbf{A}_r,\mathbf{h}_{J,k}],\forall k\in\mathcal{K}^T$ and substitute them into problem (\ref{qquestion2}). Then the transformed problem is given by
 
 \vspace{-0.3cm}
\begin{small}
 \begin{subequations}\label{qrquestion1}
 \begin{align}
 \label{qrquestiona1}
 &\max_{\mathbf{\Xi}_r,\mathbf{\Xi}_t}{\sum_{k\in\mathcal{K}^R}\log_2{\left(1+\frac{\text{Tr}(\mathbf{F}_{B,k}^H\mathbf{V}_{B,k}\mathbf{F}_{B,k}\mathbf{\Xi}_r)}{\frac{\hat{P}_J}{(1-\varepsilon_{PJ})}\text{Tr}(\mathbf{F}_{J,k}\mathbf{\Xi}_t\mathbf{F}_{J,k}^H)+\sigma^2}\right)}}\nonumber\\
 &+\sum_{k\in\mathcal{K}^T}\log_2{\left(1+\frac{\text{Tr}(\mathbf{F}_{B,k}^H\mathbf{V}_{B,k}\mathbf{F}_{B,k}\mathbf{\Xi}_t)}{\frac{\hat{P}_J}{(1-\varepsilon_{PJ})}\text{Tr}(\mathbf{F}_{J,k}\mathbf{\Xi}_r\mathbf{F}_{J,k}^H)+\sigma^2}\right)},\\
 \label{qrquestionb1}
 \text{s.t.}\ 
 &\frac{\hat{P}_J}{(1-\varepsilon_{PJ})}\text{Tr}(\mathbf{F}_{J,k}\mathbf{\Xi}_s\mathbf{F}_{J,k}^H)\le\tau_k,\forall k\in\mathcal{K},\\
  \label{qrquestionc1}
 & \text{Rank}(\mathbf{\Xi}_r)=1,\mathbf{\Xi}_r \succeq \mathbf{0},\\
 \label{qrquestiond1}
 & \mathbf{\Xi}_r(m,m)=1,\forall m \in [1,M+1],\\
  \label{qrquestione1}
  &\mathbf{\Xi}_t(m_1,m_2)=\nonumber\\
  &\left\{
     \begin{aligned}
         1,&&& m_1=m_2, m_1,m_2\in [1,M+1],\\
         e^{js\arg(\mathbf{\Xi}_r(m_1,m_2))},&&& m_1\neq m_2, m_1,m_2\in [1,M],\\
         e^{jv}e^{js\arg(\mathbf{\Xi}_r(m_1,m_2))},&&& m_1\neq m_2,m_1=M+1,\\
        e^{-jv}e^{js\arg(\mathbf{\Xi}_r(m_1,m_2))},&&& m_1\neq m_2,m_2=M+1,
     \end{aligned}
     \right.
 \end{align}
 \end{subequations}
 \end{small}
where $\mathbf{\Xi}_s=\mathbf{\Xi}_t,k\in\mathcal{K}^R$ and $\mathbf{\Xi}_s=\mathbf{\Xi}_r,k\in\mathcal{K}^T$ in constraint (\ref{qrquestionb1}). Constraint (\ref{qrquestione1}) is transformed from constraint (\ref{qquestionb2}).  

 Since the objective function (\ref{qrquestiona1}) remains non-convex, we introduce auxiliary variables $u_k$ and $t_k$ and new constraints (\ref{qrquestionc1.5}) and (\ref{qrquestiond1.5}) to make problem (\ref{qrquestion1}) convex:

 \vspace{-0.4cm}
 \begin{small}
  \begin{subequations}\label{qrquestion1.5}
 \begin{align}
 \label{qrquestiona1.5}  
&\max_{\mathbf{\Xi}_r,\mathbf{\Xi}_t,u_k,t_k}{\sum_{k\in\mathcal{K}}\log_2{(1+u_k)}},\\
 \label{qrquestionb1.5}
 ~\text{s.t.}\ &(\ref{qrquestionb1})-(\ref{qrquestione1}),\\
 \label{qrquestionc1.5}
&t_k\geq{\frac{\hat{P}_J}{(1-\varepsilon_{PJ})}\text{Tr}(\mathbf{F}_{J,k}\mathbf{\Xi}_s\mathbf{F}_{J,k}^H)+\sigma^2},\\
 \label{qrquestiond1.5}
&\frac{u_k^{(i-1)}}{2t_k^{(i-1)}}{t_k}^2+\frac{t_k^{(i-1)}}{2u_k^{(i-1)}}{u_k}^2\le\text{Tr}(\mathbf{F}_{B,k}^H\mathbf{V}_{B,k}\mathbf{F}_{B,k}\mathbf{\Xi}_s).
 \end{align}
 \end{subequations}
 \end{small}
In constraint (\ref{qrquestionc1.5}), $\mathbf{\Xi}_s=\mathbf{\Xi}_t,k\in\mathcal{K}^R$ and $\mathbf{\Xi}_s=\mathbf{\Xi}_r,k\in\mathcal{K}^T$. In constraint (\ref{qrquestiond1.5}), $\mathbf{\Xi}_s=\mathbf{\Xi}_r,k\in\mathcal{K}^R$ and $\mathbf{\Xi}_s=\mathbf{\Xi}_t,k\in\mathcal{K}^T$. The left part of constraint (\ref{qrquestiond1.5}) is an upper bound of 
$u_kt_k$, making (\ref{qrquestiond1.5}) convex and revealing the relation between $u_k$ and $t_k$. Here, we denote the current iteration number as\begin{small}
$i$\end{small}. The values of $u_k$ and $t_k$ obtained at the\begin{small}
$(i-1)$\end{small}-th iteration are denoted as $u_k^{(i-1)}$ and $t_k^{(i-1)}$, respectively. 

 \textit{iii) Channel uncertainties elimination:} Then, to eliminate the channel uncertainties $\Delta\mathbf{h}_{J,k}$ and $\Delta\mathbf{H}_{J,k}$, we introduce slack variables $\{\mathbf{\Upsilon}_k\}$ to replace the term with $\mathbf{F}_{J,k}$ in constraint (\ref{qrquestionb1}) and (\ref{qrquestionc1.5}). Constraint (\ref{qrquestionc1.5}) can be rewritten as

  \vspace{-0.3cm}
 \begin{small}
\begin{align}
\label{bk1}
t_k\geq{\frac{\hat{P}_J}{(1-\varepsilon_{PJ})}\text{Tr}(\mathbf{\Upsilon}_k)+\sigma^2},\forall k\in\mathcal{K},
\end{align}
\end{small}
and constraint (\ref{qrquestionb1}) can be rewritten as
 \begin{small}
\begin{align}
\label{C5a5}
\frac{\hat{P}_J}{(1-\varepsilon_{PJ})}\text{Tr}(\mathbf{\Upsilon}_k) \le\tau_k , \forall k\in\mathcal{K}.
\end{align}
\end{small}
Simultaneously a new constraint is introduced:

 \vspace{-0.2cm}
 \begin{small}
\begin{equation}
\begin{aligned}
\label{Upsilonk}
\mathbf{\Upsilon}_k\succeq\mathbf{F}_{J,k}\mathbf{\Xi}_s\mathbf{F}_{J,k}^H, \forall k\in\mathcal{K},
\end{aligned}
\end{equation}
\end{small}
where $\mathbf{\Xi}_s=\mathbf{\Xi}_t,k\in\mathcal{K}^R$ and $\mathbf{\Xi}_s=\mathbf{\Xi}_r,k\in\mathcal{K}^T$. The uncertainties in constraint (\ref{Upsilonk}) can be eliminated using the S-procedure~\cite{luo2004multivariate}. We define $\hat{\mathbf{F}}_{J,k}=[\hat{\mathbf{H}}_{J,k}^{H}\text{diag}(\mathbf{a}_t),\hat{\mathbf{h}}_{J,k}],\forall k\in\mathcal{K}^R$, $\hat{\mathbf{F}}_{J,k}=[\hat{\mathbf{H}}_{J,k}^{H}\text{diag}(\mathbf{a}_r),\hat{\mathbf{h}}_{J,k}],\forall k\in\mathcal{K}^T$. By applying S-procedure, constraint (\ref{Upsilonk}) can be converted into constraint (\ref{C6b}) and (\ref{C6a}) without uncertainties:

\vspace{-0.4cm}
\begin{small}
\begin{align}
\label{C6b}
\rho_{k} \geq 0, \forall k\in\mathcal{K},
\end{align}
\begin{equation}
\begin{aligned}
\label{C6a}
\begin{bmatrix} 
-\hat{\mathbf{F}}_{J,k}\mathbf{\Xi}_s\hat{\mathbf{F}}_{J,k}^H+\Upsilon_k-\rho_{k}\mathbf{I}_{N_j}\ \ \   -{\hat{\mathbf{F}}}_{J,k}\mathbf{\Xi}_s\\ 
-\mathbf{\Xi}_s{\hat{\mathbf{F}}}_{J,k}^H\ \ \ \ \ \ \ \ \ \ \ \ \ \  \frac{\rho_{k}}{{\epsilon_{d,k}}^2+{\epsilon_{J,k}}^2}\mathbf{I}_{M+1}-\mathbf{\Xi}_s
\end{bmatrix}
\succeq \mathbf{0},
\end{aligned}
\end{equation}
\end{small}
where $\mathbf{\Xi}_s=\mathbf{\Xi}_t,k\in\mathcal{K}^R$ and $\mathbf{\Xi}_s=\mathbf{\Xi}_r,k\in\mathcal{K}^T$. 

Finally, problem (\ref{qrquestion1.5}) can be reformulated as (\ref{qrquestion2}): 

 \vspace{-0.3cm}
\begin{small}
 \begin{subequations}\label{qrquestion2}
 \begin{align}
 \label{qrquestiona2}
 &\max_{\mathbf{\Xi}_r,\mathbf{\Xi}_t,\mathbf{\Upsilon}_k,\rho_{k},u_k,t_k}{\sum_{k\in\mathcal{K}}\log_2{(1+u_k)}},\\
 \label{qrquestionb2}
 ~\text{s.t.}\ & (\ref{qrquestionc1})-(\ref{qrquestione1}),(\ref{qrquestiond1.5}),(\ref{bk1}),(\ref{C5a5}),(\ref{C6b}),(\ref{C6a}).
 \end{align}
 \end{subequations}
 \end{small}

We deploy the SDR technique to remove the rank-one constraint in (\ref{qrquestionc1}) and allow the problem to be solved with convex solvers. If the obtained $\mathbf{\Xi}_r$ satisfies $\text{Rank}(\mathbf{\Xi}_r)=1$, $\mathbf{q}_r$ can be recovered by the eigenvalue decomposition. If not, a rank-one solution is typically constructed by the Gaussian randomization technique~\cite{luo2010semidefinite}. Then $\{\phi_m^r\}$ can be obtained from $\mathbf{q}_r$, and $\{\phi_m^t\}$ can be obtained from constraint (\ref{qquestionb2}). The obtained continuous reflection and refraction phase shifts are denoted by $\{\phi_m^{r,c}\}$ and $\{\phi_m^{t,c}\}$, respectively.

\subsubsection{Step 3 - Discrete and coupled IOS phase shift design}
However, $(\phi_m^{r,c},\phi_m^{t,c})$ may not correspond to any option in $\mathcal{F}$. In general, the continuous phase shift of the $m$-th element falls between two discrete phase shifts in $\mathcal{F}$, i.e., 
$\varphi_{l_m}^r\le\phi_m^{r,c}<\varphi_{l_m+1}^r$, $\varphi_{l_m}^t\le\phi_m^{t,c}<\varphi_{l_m+1}^t$. 
Therefore, the discrete phase shifts of the $m$-th element $(\phi_m^r,\phi_m^t)$ is selected from $\mathcal{F}_m=\{(\varphi_{l_m}^r,\varphi_{l_m}^t),(\varphi_{l_m+1}^r,\varphi_{l_m+1}^t)\}$. To reduce complexity, we use the local search method. First, we set each element's phase shifts to their continuous values from step~2. Then, fixing the phase shifts of the other $M-1$ elements, we iterate through each element's phase shift options in $\mathcal{F}_m$ and select the one that maximizes the sum rate and satisfies the jamming constraint. We use this option for the $m$-th element and continue optimizing other elements until all are optimized. Algorithm~1 summarizes the IOS phase shift design.

\begin{algorithm}
\begin{small}
\DontPrintSemicolon
  \SetAlgoLined
   \KwIn {Digital beamforming $\mathbf{V}_B$.}
   \KwOut {IOS reflection phase shifts $\{\phi_m^r\}$ and refraction phase shifts $\{\phi_m^t\}$.}
  \SetKwProg{Fna}{Step 1:}{}{End}
  \SetKwProg{Fn}{Step 2:}{}{End}
  \SetKwProg{Fnb}{Step 3:}{}{End}
  Initialize $u_k^{(0)}$, $t_k^{(0)}$, $\mathbf{\Upsilon}_{k}^{(0)}$, and iteration number $i=1$.\;
    \Repeat{convergence.}{
      Obtain $u_k$, $t_k$, $\mathbf{\Upsilon}_{k}$ and $\mathbf{\Xi}_r$ by solving (\ref{qrquestion2}).\;
      Update variables $u_k^{(i)} \gets u_k$, $t_k^{(i)} \gets t_k$, $\mathbf{\Upsilon}_{k}^{(i)}  \gets \mathbf{\Upsilon}_{k}$;
      Update $i \gets i+1$.\;
    } 
    Obtain $\{\phi_m^{r,c}\}$ from $\mathbf{\Xi}_r$. Obtain $\{\phi_m^{t,c}\}$ from $\{\phi_m^{r,c}\}$.\;
    Set $\phi_m^r=\phi_m^{r,c}$ and $\phi_m^t=\phi_m^{t,c}$ for all $M$ elements.\;
    \For{m=1:M}{
     Iterate all options in $\mathcal{F}_m$ to $\phi_m^{r}$ and $\phi_m^{t}$. Select the one maximizing the sum rate on the premise that constraint (\ref{qquestione}) is satisfied.
  }
  \caption{IOS Phase Shift Design}
  \end{small}
\end{algorithm}


\subsection{Overall Algorithm Description}
Based on the previous two subsections, we propose an overall algorithm for solving the original problem (\ref{sequestion}) in an iterative manner. Specifically, we initialize IOS phase shifts first and then solve the two subproblems alternately. In the $n$-th iteration, we first optimize the digital beamformer $\mathbf{V}_B$ given the IOS phase shifts as described in Section~\ref{vbopt}. Then, given the digital beamformer $\mathbf{V}_B$, the IOS reflection phase shifts $\{\phi_m^r\}$ and refraction phase shifts $\{\phi_m^t\}$ are optimized by using Algorithm~1. Those obtained $\{\phi_m^r\}$ and $\{\phi_m^t\}$ are set as the initial solutions for the next iteration.  The iteration continues until the sum rate converges. 


\vspace{-0.2cm}
 \begin{figure*}[!t]
 \subfigure[]{
\begin{minipage}[t]{0.32\linewidth}
\centering
\includegraphics[width=0.9\textwidth,height=1.65in]{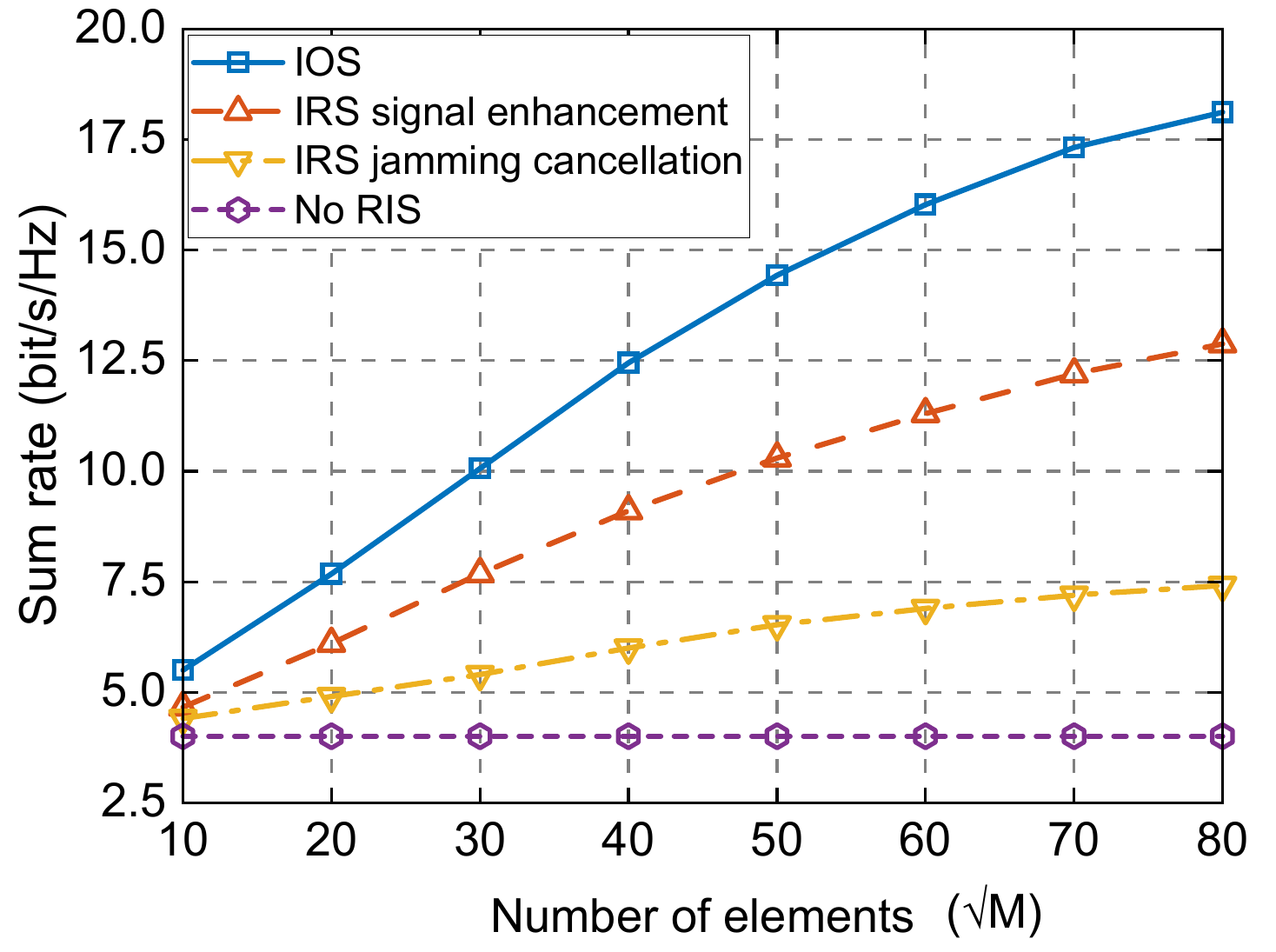}
\label{sumrate_number}
\end{minipage}%
}
\subfigure[]{
\begin{minipage}[t]{0.32\linewidth}
\centering
\includegraphics[width=0.9\textwidth,height=1.65in]{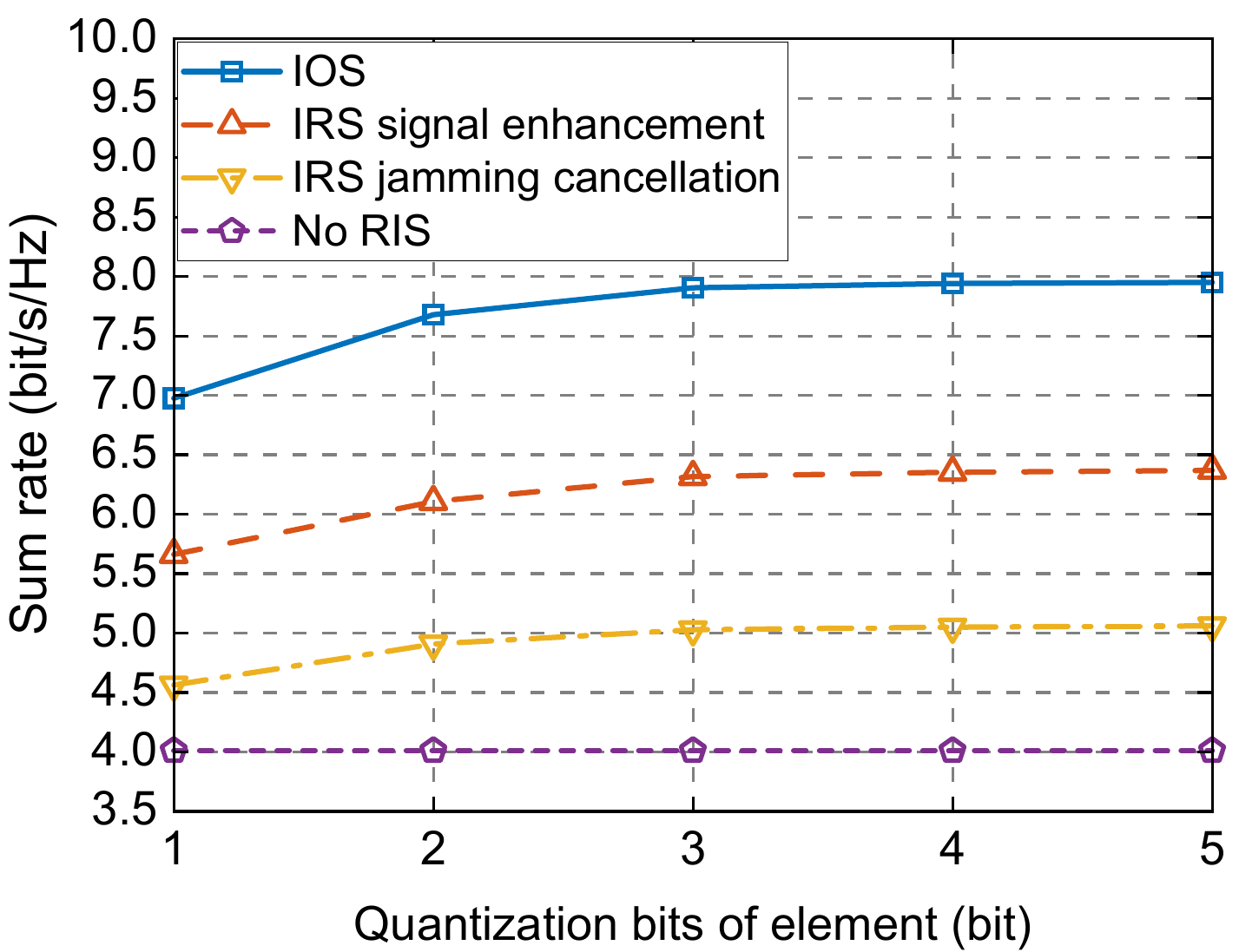}
\label{sumrate_bit}
\end{minipage}
}
\subfigure[]{
\begin{minipage}[t]{0.32\linewidth}
\centering
\includegraphics[width=0.9\textwidth,height=1.65in]{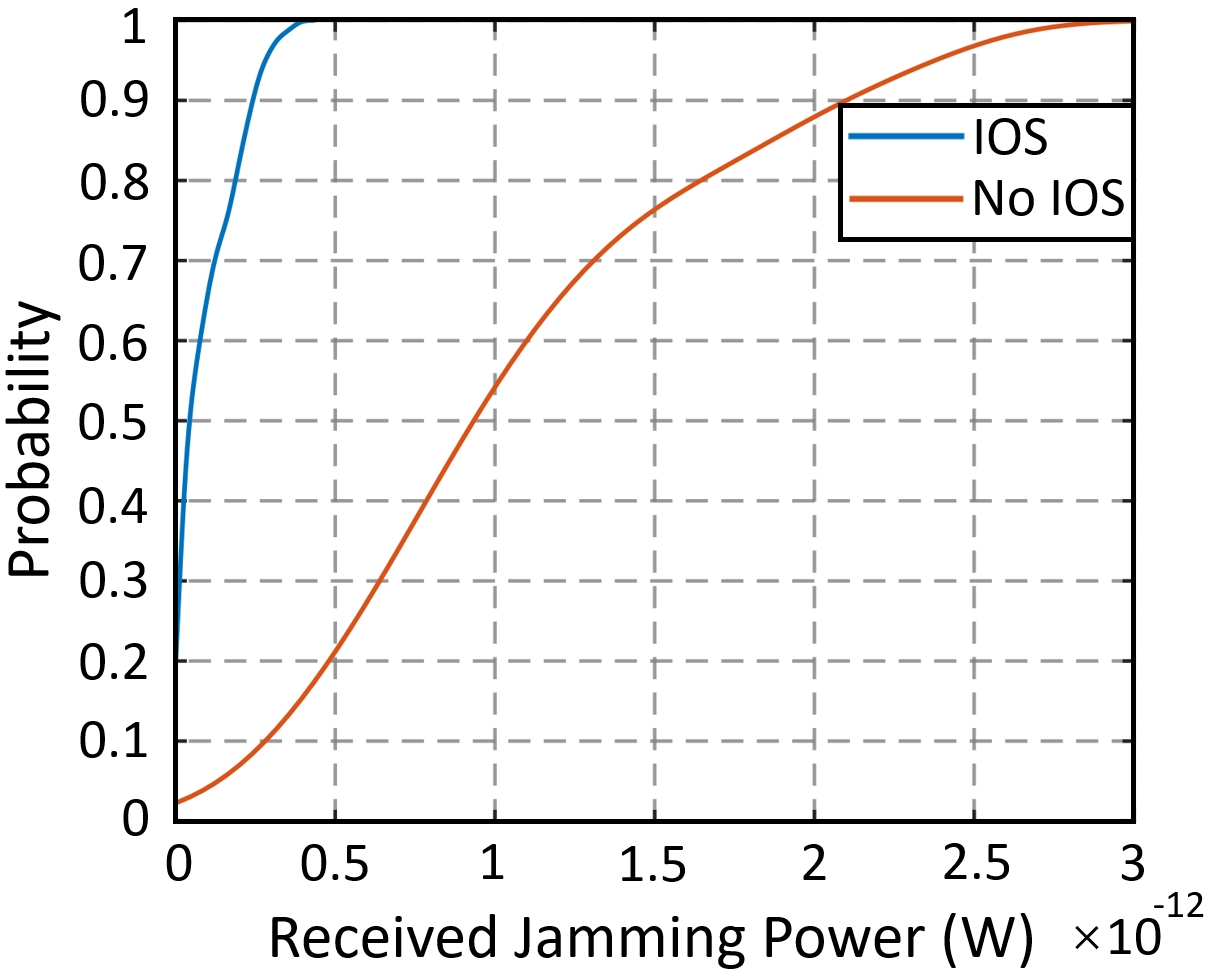}
\label{ganrao}
\end{minipage}
}
\vspace{-0.6cm}
\caption{(a) surface size vs. sum rate ($b=2$). (b) quantization bits vs. sum rate ($M=400$). (c) Cumulative distribution function of user-received jamming power with and without IOS.}
\vspace{-0.6cm}
\end{figure*}

 \section{Simulation Results}
In this section, we assess the performance of the proposed IOS-aided anti-jamming scheme. In the simulations, both the BS and the jammer are 100 m from the IOS, with users randomly deployed within a disk of radius 50 m centered at the IOS. The total transmit power of the BS and the jammer is 40 dBm, the noise power is -96 dBm, and the carrier frequency is 28 GHz. The number of BS antennas and jammer's antennas are set to $N_b = 32$ and $N_j = 8$ as in~\cite{yang2020intelligent} and~\cite{sun2021intelligent}, respectively. The antenna separation is half the wavelength, and the inter-distance between IOS elements is 0.005 m. The power ratio of the reflected and refracted signals is $\epsilon= 1$. The path-loss exponent of the direct link is 3, and the Rician factor is $\kappa= 4$~\cite{zhang2021intelligent}. The jammer's beamforming is $\mathbf{v}_J=\sqrt{P_J}(\hat{\mathbf{h}}_{J,k_{near}}/\Vert\hat{\mathbf{h}}_{J,k_{near}}\Vert_2)$~\cite{sun2021intelligent}, where $k_{near}$ denotes the nearest user to jammer. Power estimation error is $\varepsilon_{PJ}=0.1$~\cite{sun2022outage}. We define $\zeta_{d,k}$ and $\zeta_{J,k}$ as the maximum normalized estimation error for the channel $\mathbf{h}_{J,k}$ and $\mathbf{H}_{J,k}$, i.e., $\zeta_{d,k}=\epsilon_{d,k}/\Vert\hat{\mathbf{h}}_{J,k}\Vert_2$ and $\zeta_{J,k}=\epsilon_{J,k}/\Vert\hat{\mathbf{H}}_{J,k}\Vert_F$, respectively, and fix $\zeta_{d,k}^2=\zeta_{J,k}^2=0.1$. The maximum received jamming power of users is $\tau_k=0.3\tau_{d,k}$, where $\tau_{d,k}$ is the jamming power received by user $k$ without any RIS.

Three schemes are performed for comparison. \textit{1)IRS-assisted signal enhancement:} The IRS faces the BS, reflecting only the desired signal from the BS to the users on one side for enhancement. \textit{2)IRS-assisted jamming cancellation:} The IRS faces the jammer, reflecting only the jamming signal from the jammer to the users on one side for cancellation. \textit{3)No RIS:} All users receive the desired signal and the jamming signal through direct links.

Fig.~\ref{sumrate_number} shows the sum rate as a function of surface size. The sum rate increases with element numbers, and the growth rate decreases with element numbers. Within the exhibited range of element numbers, the IOS scheme achieves a higher sum rate than other schemes since it simultaneously enhances the desired signal and nullifies the jamming for all users. 

Fig.~\ref{sumrate_bit} displays the sum rate as a function of the quantization bits $b$ for each IOS or IRS element. The sum rate increases with $b$ and converges as $b$ continues to increase, with 3 bits providing a near-optimal sum rate. Within the range of 1 to 5 bits, the IOS scheme outperforms other schemes. Notably, a 1-bit IOS achieves a higher sum rate compared to a high-bit IRS, showing the advantages of the IOS scheme. 

Moreover, Fig.~\ref{ganrao} presents the jamming cancellation effects achieved by the IOS scheme. Compared to not using IOS, there is a significant leftward shift in the cumulative distribution function of the user-received jamming power with the IOS scheme, where 94$\%$ of users experience a reduction in received jamming power to below -96 dBm. This indicates that the IOS scheme exhibits effective jamming cancellation.

 \section{Conclusion}
 \vspace{-0.1cm}
In this paper, we have considered an IOS-aided multi-user anti-jamming wireless communication system where the IOS is used to simultaneously nullify jamming and enhance desired signals by utilizing its reflection and refraction properties. We have developed a joint BS digital beamforming and IOS analog beamforming algorithm to maximize the sum rate under the tolerable jamming constraint. To deal with the coupled and discrete IOS phase shifts, we relax them to continuous states and optimize using a coupling-aware algorithm, followed by a local search to recover discrete states. Based on simulations, three main conclusions can be drawn:
\begin{enumerate}
\item{The proposed scheme has the capability of simultaneous jamming cancellation and desired signal enhancement.
}
\item{The IOS scheme achieves a higher sum rate compared to IRS schemes in the presence of jamming attacks. 
}
\item{Using a 1-bit IOS achieves a higher sum rate compared to using a high-bit IRS.}
\end{enumerate}

\vspace{-0.1cm}
\section*{Acknowledgement}
\vspace{-0.1cm}
This work was supported in part by the National Science Foundation under Grants 62271012, 62371011, and 62401024; in part by the Beijing Natural Science Foundation under Grant L243002; in part by the GuangDong Basic and Applied Basic Research Foundation under Grant 2023B0303000019; and in part by the Pengcheng Laboratory Major Key Project under Grant  PCL2024A01.


\end{document}